\documentclass[aps,twocolumn]{revtex4-1}
\usepackage{newlfont}
\usepackage{amssymb}
\usepackage{amsfonts}
\usepackage{amsmath}
\usepackage{wasysym}
\usepackage{graphicx}
\usepackage{bm}

\usepackage{graphicx}
\usepackage{epsfig}
\usepackage{newlfont}
\usepackage{amssymb}
\usepackage{amsfonts}
\usepackage{amsmath}
\usepackage{graphicx}
\usepackage{bm}

\usepackage{amsthm}

\begin{document}

\title{Characterization of Tripartite Quantum States with Vanishing Monogamy Score}


\author{Manabendra N. Bera, R. Prabhu, Aditi Sen(De), and Ujjwal Sen}

\affiliation{Harish-Chandra Research Institute, Chhatnag Road, Jhunsi, Allahabad 211 019, India}



\begin{abstract}

Quantum discord, an information-theoretic quantum correlation measure, can satisfy as well as violate monogamy, for three-party quantum
states. We quantify the feature using  the concept of discord monogamy score.  We find a necessary condition of a vanishing discord monogamy score for arbitrary three-party states. A necessary and sufficient condition is obtained for pure states. We prove that the class of states having a vanishing discord monogamy score cannot have arbitrarily high genuine multipartite entanglement, as quantified by generalized geometric measure. In the special case of three-qubit pure states, their classification with respect to the discord monogamy score, reveals a rich structure that is different from that which had been obtained by using the monogamy score corresponding to the entanglement measure called concurrence. We investigate properties like genuine multipartite entanglement and violation of the multipartite Bell inequality for  these states.

\end{abstract}

\maketitle

\section{Introduction}

The concept of monogamy is an important binding theme for  quantum correlations \cite{HHHH-RMP} of states 
shared between several quantum systems. 
Monogamy arises due to the fact that bipartite quantum correlations of states of three or more quantum systems are usually such that if two of the parties 
are highly quantum correlated, these parties have little or no quantum correlations with any other party \cite{Ekert91, Bennett-prathham, 
Wootters, KW04, monogamyN}. One can assign quantum monogamy scores corresponding to multiparty quantum systems, to quantify 
their monogamous nature, or a possible violation thereof \cite{Wootters, acc-monogamy, discord-monogamy, aalo}. 
Such scores would typically depend on the quantum correlation measure used, and while 
the entanglement measure, concurrence squared \cite{Hill-Wootters},  
was considered in Ref. \cite{Wootters}, the information-theoretic quantum correlation measure,
quantum discord \cite{discord}, was studied in Refs. \cite{discord-monogamy, aalo}. 

In the case when concurrence squared is used as a measure of quantum correlations to construct the quantum monogamy score, the class 
of states with  vanishing scores was identified in Ref. \cite{Wootters}, for the case of 
three-qubit pure states. They constitute the important family of W-class states \cite{Wstate, dur-vidal-cirac}, complementary 
to the Greenberger-Horne-Zeilinger (GHZ)-class ones \cite{GHZ, dur-vidal-cirac}. 
We consider the quantum monogamy score, with the quantum correlation measure being quantum discord, and call it 
 the ``discord monogamy score''. 
For  arbitrary tripartite quantum states, we characterize the class possessing a vanishing discord monogamy score. 
We find necessary and sufficient conditions for tripartite quantum states with zero discord monogamy score. 
More specific results are obtained for 
three-qubit pure states. In particular, while none of the genuinely multiparty entangled W-class states have a vanishing discord monogamy score ( in stark contrast to the 
case when concurrence is used as the quantum correlation measure to build the quantum monogamy score), there does exist genuinely multiparty entangled GHZ-class states having a zero discord monogamy score.
We characterize the pure three-qubit states with zero discord monogamy score, as well as those with  negative and positive values of the same, by using a measure of genuine multiparty entanglement, called generalized geometric measure (GGM) 
\cite{amaderGGM}, and 
by using the Mermin-Klyshko Bell inequalities \cite{MK}. Specifically, we find that states with a vanishing discord monogamy score can have substantially high values of genuine multisite entanglement. Subsequently, however, we show that such states 
cannot have a maximal GGM. We also find a relation between the bipartite entanglements of formation of a tripartite three-qubit state and its generalized geometric measure. 
It may be noted here that the discord monogamy score has been interpreted as a multiparty information-theoretic quantum correlation measure \cite{aalo}, just like the quantum monogamy score with respect to 
concurrence squared has been reasoned as a multiparty entanglement measure \cite{Wootters}.

Below, in Sec. \ref{raat-ekhon-poune-char}, 
we define the measures of bipartite quantum correlations that we use later in the paper. They are 
respectively the entanglement of formation, the concurrence, and the quantum discord. 
We subsequently define the quantum monogamy scores in Sec. \ref{poali-r-khale-switch-tepa-chhata}.
The results are presented in the following section, viz. Sec. \ref{lola-lolu}, and in Sec. \ref{air-cooler-er-jol-sesh}.
The considerations in Sec. \ref{air-cooler-er-jol-sesh} require the introduction of a genuine multiparty entanglement measure, and the Mermin-Klyshko inequalities.
The generalized geometric measure is defined as a measure of genuine multisite entanglement in Sec. \ref{khub-mosha}, and  
the Mermin-Klyshko Bell inequalities are discussed in Sec. \ref{ghum-peye-gyachhe}.
We present a conclusion in Sec. \ref{poune-paanch}.

\section{Bipartite quantum correlations}
\label{raat-ekhon-poune-char}

In this section, we define quantum correlation measures that are thereafter employed in the succeeding 
 sections to obtain the corresponding quantum  monogamy scores.
These quantum monogamy scores will be helpful in formulating a classification scheme for tripartite quantum states.

\subsection{Entanglement of formation}

The entanglement of formation of a bipartite quantum state is the amount of singlets
required to prepare a state by local quantum operations and classical communication. 
If \(|\psi\rangle_{AB}\) is a bipartite state shared between two parties \(A\) and \(B\), then it can be shown that entanglement of formation is equal to the von Neumann entropy of its local density matrix
\cite{Bennett-prathham,Bennett-motka-paper}: 
\begin{equation}
E^f(|\psi\rangle_{AB})= S(\varrho_A) = S(\varrho_B).
\end{equation} 
where $\varrho_{A}$ and \(\varrho_B\) are the local density matrices of the combined system $|\psi\rangle_{AB}$, obtained by performing partial traces over subsystems $B$ and $A$ respectively, and 
\(S(\sigma) = - \mbox{tr} \left(\sigma \log_2 \sigma\right)\) is the von Neumann entropy of a quantum state \(\sigma\). 
Entanglement of formation of a mixed bipartite state \(\rho_{AB}\) is 
then defined by the convex-roof approach: 
\begin{equation}
E^f(\rho_{AB})=\mbox{min}\sum_i p_iE^f(|\psi_i\rangle_{AB}),
\label{mixeof}
\end{equation}
where the minimization is over all pure state decompositions of $\rho_{AB} = \sum_i p_i (|\psi_i\rangle \langle \psi_i|)_{AB}$.
We often denote \(E^f(\rho_{AB})\) simply as \(E^f_{AB}\). 


\subsection{Concurrence}
\label{sec:concurrence}

The concept of concurrence is derived from the definition of entanglement of formation and is proposed to quantify the entanglement 
of two-qubit  states \cite{Hill-Wootters}. The definition of entanglement of formation for mixed states (see Eq. (\ref{mixeof})) involves a minimization which is in general not easy to perform. 
However, there exists a closed form for the case of two-qubit states \cite{Hill-Wootters}, in terms of the concurrence, which is 
defined as $C(\rho_{AB})=\mbox{max}\{0,\lambda_1-\lambda_2-\lambda_3-\lambda_4\}$, and often denoted below as \(C_{AB}\).  Here the $\lambda_i$'s are the square roots of the eigenvalues of $\rho\tilde{\rho}$ in decreasing order and 
$\tilde{\rho}=(\sigma_y\otimes\sigma_y)\rho^*(\sigma_y\otimes\sigma_y)$, with the complex conjugation being
in the computational basis. $\sigma_y$ is the Pauli spin matrix.

\subsection{Quantum discord}
\label{sec:discord}

In classical information theory, there are two equivalent ways to define the mutual information between two random variables. One of the methods is 
 by adding the Shannon entropies of the individual random variables, and then subtracting the same of 
the joint probability distribution.  
%
 The second one is to use the concept of conditional entropies. 
%
%

Quantizing these two classically equivalent definitions of mutual information  gives rise to two inequivalent concepts, whose difference is termed as quantum discord \cite{discord}. 
Quantizing the former classical definition of mutual information is performed by replacing the Shannon entropies by von Neumann entropies: For a quantum state \(\rho_{AB}\) of two parties, the ``quantum mutual information'' 
is defined as \cite{qmi} (see also \cite{Cerf, GROIS})
\begin{equation}
\label{qmi1}
I(\rho_{AB})= S(\rho_A)+ S(\rho_B)- S(\rho_{AB}),
\end{equation}
where \(\rho_A\) and \(\rho_B\) are the local density matrices of \(\rho_{AB}\).

Quantizing the latter classical definition of mutual information is more involved, as replacing the Shannon entropies by von Neumann ones in this case will lead to a  physical quantity which can assume negative values for some quantum states \cite{Cerf}.
Such a shortcoming is  overcome by  
interpreting the conditional entropy in the classical case as a measure of the lack of information about one of the random variable, 
when the other is known in a joint probability distribution of two random variables.
This leads to the following quantization of the classical mutual information  for 
a bipartite quantum state \(\rho_{AB}\):
\begin{equation}
\label{cmi1}
 J(\rho_{AB}) = S(\rho_A) - S(\rho_{A|B}),
\end{equation}
where the ``quantum conditional entropy'', \(S(\rho_{A|B})\equiv S_{A|B}\), is defined as  
\begin{equation}
\label{qce}
S(\rho_{A|B}) = \min_{\{\Pi_i^B\}} \sum_i p_i S(\rho_{A|i}),
 \end{equation}
with the minimization being over all 
rank-1
measurements, \(\{\Pi^B_i\}\),  performed on subsystem \(B\).
Here \(p_i = \mbox{tr}_{AB}(\mathbb{I}_A \otimes \Pi^B_i \rho_{AB} \mathbb{I}_A \otimes \Pi^B_i)\) is the probability for obtaining the outcome \(i\), and 
the corresponding post-measurement state 
for the subsystem \(A\) is \(\rho_{A|i} = \frac{1}{p_i} \mbox{tr}_B(\mathbb{I}_A \otimes \Pi^B_i \rho_{AB} \mathbb{I}_A \otimes \Pi^B_i)\), 
where \(\mathbb{I}_A\) is the identity operator on the Hilbert space of the quantum system that is in possession of \(A\).



The difference between these two inequivalent quantized version of the classical mutual information
is termed as quantum discord. 
Also, it has be established that the quantum mutual information is never lower than the quantity \(J\). Therefore, the 
quantum discord is given by 
\cite{discord}
\begin{equation}
D(\rho_{AB})= I(\rho_{AB}) - J(\rho_{AB}).
\end{equation}
Unlike  many other measures of quantum correlations, even separable states may produce a nonzero discord.
We often denote \(D(\rho_{AB})\) as \(D_{AB}\).

\section{Quantum Monogamy scores}
\label{poali-r-khale-switch-tepa-chhata}

The sharing of quantum correlations among subsystems of a multiparticle quantum state is often constrained by the concept of monogamy. In the tripartite scenario, 
if $\cal{Q}$ is a bipartite quantum correlation measure, then this  measure is said to be monogamous (or satisfy monogamy) for a tripartite quantum state \(\rho_{ABC}\), and with \(A\) as the ``nodal observer'', if
\begin{eqnarray}
\cal{Q}(\rho_{A:BC}) \geq \cal{Q}(\rho_{AB}) +  \cal{Q}(\rho_{AC}).  
\end{eqnarray}
Here $\cal{Q}(\rho_{AB})$ is the quantum correlation (with respect to the measure \(\mathcal{Q}\)) between subsystems $A$ and $B$, $\cal{Q}(\rho_{AC})$ is the quantum correlation between subsystems $A$ and $C$, and $\cal{Q}(\rho_{A:BC})$ is 
quantum correlation between subsystem $A$ and subsystems $B$ and $C$ taken together. 
By rearranging the terms in the above inequality, we get
\begin{eqnarray}
\delta_\cal{Q} \equiv \cal{Q}(\rho_{A:BC}) - \cal{Q}(\rho_{AB}) -  \cal{Q}(\rho_{AC}) \geq 0.
\end{eqnarray}
This leads to the concept of quantum monogamy score, which, for a given bipartite quantum correlation measure, is defined as 
\begin{eqnarray}
\delta_\cal{Q} \equiv \cal{Q}(\rho_{A:BC}) - \cal{Q}(\rho_{AB}) -  \cal{Q}(\rho_{AC}),
\end{eqnarray}
 irrespective of whether it monogamous or not. 

In Ref. \cite{Wootters}, concurrence squared was used as the quantum correlation measure to define a quantum monogamy score. We denote it by \(\delta_C\), and call it as entanglement monogamy score. 
Discord monogamy score was introduced in Refs. \cite{discord-monogamy, aalo}, and is the quantum monogamy score when one uses the quantum discord as the quantum correlation measure. We denote it as \(\delta_D\), and it is defined, with 
\(A\) as the nodal observer, as 
\begin{eqnarray}
\delta_D = D(\rho_{A:BC}) - D(\rho_{AB}) -  D(\rho_{AC}).  
\label{eq:dismono}
\end{eqnarray}
 Interestingly, these two quantum correlation measures (concurrence and quantum discord) have been studied together and their opposing statistical mechanical behaviors are reported in  Ref. \cite{Ergodiscord}.



\section{Conditions for vanishing discord monogamy score}
\label{lola-lolu}

In this section, we provide some general features that are exhibited by states of vanishing discord monogamy score. 
Consider a three-party quantum state, \(\rho_{ABC}\), which can be pure or mixed, in arbitrary dimensions.


\noindent \textbf{Proposition I:} For an arbitrary quantum state \(\rho_{ABC}\), a necessary condition for discord monogamy score to be vanishing, with \(A\) as the nodal observer,  is given by  
\begin{eqnarray}
\mathcal{D}_{A:BC} \leq S_{A|B} + S_{A|C}.  
\label{eq:condi}
\end{eqnarray}
\\

\noindent \texttt{Proof.}
A vanishing discord monogamy score implies 
\begin{equation}
 2S_A - \mathcal{D}_{A:BC} = -S_B-S_C+S_{AB} + S_{AC} + J_{AB} + J_{AC},
\label{moutusi}
\end{equation}
where \(S_A\) denotes the von Neumann entropy of \(\mbox{tr}_{BC}(\rho_{ABC})\), and similarly for the other von Neumann entropies. 
Strong subadditivity of von Neumann entropy implies 
\begin{equation}
-S_B-S_C+S_{AB} + S_{AC} \geq 0,
\end{equation}
which leads us to  
\begin{equation}
 2S_A - \mathcal{D}_{A:BC}  \geq J_{AB} + J_{AC},
\end{equation}
which can be further simplified to obtain the stated result. \hfill \(\blacksquare\)

It may be interesting to note that the vanishing of discord monogamy score is equivalent to the statement that 
\begin{equation}
 I_{ABC} = J_{AB} + J_{AC},
\end{equation}
where 
\begin{equation}
 I_{ABC} = S_A +S_B + S_C -S_{AB} - S_{BC} - S_{CA} + S_{ABC}
\end{equation}
is the tripartite quantum interaction information \cite{interaction-info, acc-monogamy, discord-monogamy, aalo}.

For pure tripartite states, it is possible to obtain a necessary and sufficient condition for vanishing of discord monogamy score. 

\noindent \textbf{Proposition II:} For an arbitrary pure quantum state \(|\psi_{ABC}\rangle\), a necessary and sufficient condition for vanishing discord monogamy score, with \(A\) as the nodal observer,
is 
\begin{eqnarray}
S_A = S_{A|B} + S_{A|C}.  
\label{eq:condi-pure}
\end{eqnarray}
\\

\noindent \texttt{Proof.} A vanishing discord monogamy score again implies Eq. (\ref{moutusi}). However, for pure three-party states,  
\begin{equation}
-S_B-S_C+S_{AB} + S_{AC} = 0.
\end{equation}
Further, quantum discord for a pure state is just the local von Neumann entropy. Hence, the proof. \hfill \(\blacksquare\)


Interestingly, for symmetric pure tripartite states, the necessary and sufficient condition reduces to 
\begin{eqnarray}
\frac{1}{2} S_A = S_{A|B}.
\label{eq:condi-joruri}
\end{eqnarray}

From the propositions above, it is clear that the vanishing of discord monogamy score is intimately related to the sum, \(S_{A|B}+S_{A|C}\), of the quantum conditional entropies. This is particularly true for pure three-party states. 
It is therefore important to have estimates of this quantity, as finding it in closed form may sometimes be difficult. 


To obtain the bounds, let us first note that in the case of pure three-party states, we have  \cite{KW04} 
\begin{equation}
 E^f_{AB} + J_{AC} = S_A,
\end{equation}
so that we get
\begin{equation}
{\cal D}_{AB} + {\cal D}_{AC} = E^f_{AB} +  E^f_{AC}.
\end{equation}
This relation can now be used to obtain estimates on the sum of quantum conditional entropies. In particular, 
using \cite{Bennett-motka-paper}
\begin{eqnarray}
E^f_{AB} \leq \min[S_A , S_B ],\\
E^f_{AC} \leq \min[S_A, S_C],
\label{eq:EoFupperbound}
\end{eqnarray}
we get an upper bound as 
\begin{equation}
S_{A|B} + S_{A|C} \leq \min[S_A , S_B ] + \min[S_A , S_C ].
\end{equation}

On the other hand, to obtain a lower bound, we  use the lower bound
\cite{Leibrecent}
\begin{eqnarray}
E^f_{AB} \geq \max[S_A - S_{AB}, S_B - S_{AB}, 0]\\
E^f_{AC} \geq \max[S_A - S_{AC}, S_C - S_{AC}, 0],
\label{eq:EoFlowerbound}
\end{eqnarray}
to have
\begin{eqnarray}
\max[S_A - S_{AB}, S_B - S_{AB}, 0]   \phantom{aaaaaaaaaaaaaaaaaa}   \nonumber \\
+ \max[S_A - S_{AC}, S_C - S_{AC}, 0]  \leq S_{A|B} + S_{A|C}.
\label{eq:condipure}
\end{eqnarray}

\section{Generalized Geometric Measure}
\label{khub-mosha}

A multiparty pure quantum state is said to be \emph{genuinely} multiparty entangled, if it is entangled across all 
bi-partitions of its constituent parties. Quantification of genuine multiparty entanglement in such systems can be obtained by using the generalized geometric measure introduced in Ref. \cite{amaderGGM}.
%
The GGM of  an \(N\)-party pure quantum state \(|\phi_N\rangle\) is defined as
\begin{equation}
{\cal E} ( |\phi_N\rangle ) = 1 - \Lambda^2_{\max} (|\phi_N\rangle ), 
\end{equation}
where  \(\Lambda_{\max} (|\phi_N\rangle ) =
\max | \langle \chi|\phi_N\rangle |\), with  the maximization being over all pure states \(|\chi\rangle\)
that are not genuinely \(N\)-party entangled. 
It is shown in Ref. \cite{amaderGGM} that 
\begin{equation}
\label{label}
{\cal E} (|\phi_N \rangle ) =  1 - \max \{\lambda^2_{{\cal A}: {\cal B}} |  {\cal A} \cup  {\cal B} = 
\{1,2,\ldots, N\},  {\cal A} \cap  {\cal B} = \emptyset\},
\end{equation}
where \(\lambda_{{\cal A}:{\cal B}}\) is  the maximal Schmidt coefficient in the \({\cal A}: {\cal B}\) 
bipartite split  of \(|\phi_N \rangle\).

\section{Mermin-Klyshko Bell inequalities}
\label{ghum-peye-gyachhe}

Bell inequalities are relations that are derived to satisfy any physical theory that is consistent with local realism \cite{Bellbook}. Quantum mechanics is known to 
violate such inequalities. There are a large number of such inequalities known in the multiparty domain, and we will 
be using the ones which have been called the 
Mermin-Klyshko (MK) Bell inequalities \cite{MK}. A Bell operator for the Mermin-Klyshko inequalities for \(N\) qubits,
can be defined recursively as \cite{ScaraniGisinBelloperator}
\begin{equation}
\label{MK}
 B_{k} = \frac{1}{2} B_{k-1} \otimes ( \sigma _{a_k} + \sigma _{a_k^{'}})
 + \frac{1}{2} B^{'}_{k-1} \otimes ( \sigma _{a_k} - \sigma _{a_k^{'}}),
\end{equation}
where \(B^{'}_{k}\) are obtained from \( B_{k}\) by interchanging
\(a_{k}\) and \(a_k^{'}\),
and
\[B_1 = \sigma_{a_1} \quad \mbox{and} \quad B_{1}^{'} = \sigma_{a_{1}^{'}}.\]
The party \(A_j\) is allowed to choose between
the measurements \(\sigma_{a_j}\) and \(\sigma_{a^{'}_j}\), \(a_j\)
and \(a^{'}_{j}\)
being two three-dimensional unit vectors (\(j = 1, 2, \ldots, N\)).

An \(N\)-qubit state \(\eta\)  is said
to violate the
 MK inequality, and hence violate local realism, if
\begin{equation}
\label{eq_criterion_MK}
\left|\mbox{tr}\left(B_{N}\eta\right)\right| > 1.
\end{equation}


\section{Three-qubit systems}
\label{air-cooler-er-jol-sesh}

In this section, we find further properties of states with zero discord monogamy score, where we specialize to the case of three-qubit pure states. Genuinely entangled three-qubit pure 
states are known to have an important classification in terms of transformability under stochastic local quantum operations and classical communication. They form respectively the GHZ and the W classes \cite{dur-vidal-cirac}. The concept 
of quantum monogamy score, where concurrence squared is used as the quantum correlation measure \cite{Wootters}, also leads to a classification. However, these classifications are one and the same: W-class states are those for which 
the entanglement monogamy score is zero, and GHZ-class states are those for which the same is positive, there being no three-qubit pure states with a negative entanglement monogamy score \cite{Wootters}. Discord monogamy score however leads to a 
richer structure in the same space. First of all, there are pure three-qubit states that have negative, zero, and positive scores. Moreover, the negative region is occupied by both states from the GHZ and W classes, 
with the positive region covered by GHZ-class states only. Also,
there are no genuinely tripartite entangled states of the W-class that have zero discord monogamy score \cite{discord-monogamy, aalo, Giorgi}.

Tripartite pure states that are not genuinely multiparty entangled, certainly have a vanishing discord monogamy score. This is because such states are bi-separable (or maybe even tri-separable), and so are of the form 
%
%
\(|\psi_A^1\rangle \otimes |\psi^2_{BC}\rangle\) or  \(|\psi^1_B\rangle \otimes |\psi^2_{AC}\rangle\) or \(|\psi^1_C\rangle \otimes |\psi^2_{CA}\rangle\). Therefore, \(\delta_D = 0\) with any of the options, and with 
any observer as the nodal observer. 

Therefore, barring the cases when there is no genuine tripartite entanglement, all three-qubit pure states having zero discord monogamy score, are within the GHZ class. 
An arbitrary GHZ class state (unnormalized) can be written as \cite{dur-vidal-cirac}
\begin{equation}
\label{radio-churi-jabar-bedanai-amar-bandhu-Jagadis}
 |\psi_{GHZ} \rangle = \cos \theta |000\rangle + \exp (i \kappa) \sin \theta |\phi_1 \phi_2 \phi_3\rangle, 
\end{equation}
up to local unitaries, where \(|\phi_{j}\rangle = \cos\alpha_j |0\rangle + \sin\alpha_j |1\rangle\) \((j=1,2, 3)\). Here \(\theta \in (0,\pi/4]\), \(\kappa \in [0,2\pi]\), and \(\alpha_j \in (0,\pi/2]\).
An important family of states within the GHZ class  are those for which the \(|\phi_j\rangle\)s are equal: they are symmetric GHZ-class states. In that case, let 
\(\alpha_1 = \alpha_2 = \alpha_3 = \alpha\), and let the corresponding states be denoted by \(|\psi_{GHZ}^s\rangle\).
Let us begin by considering the explicit equation characterizing the \(\delta_D=0\) states within this class of symmetric states.
The concurrence of any of the two-party states, obtained by tracing out one party from \(|\psi^s_{GHZ}\rangle\),
is given by \cite{Hill-Wootters}
\begin{eqnarray}
 \label{eq:sym_GHZ2}
{\cal C}= \sqrt{\lambda_1} - \sqrt{\lambda_2},
\end{eqnarray}
where 
\begin {eqnarray}
\lambda_1 & = & (a + b)c \nonumber\\
\lambda_2 & = & (a - b)c
\end {eqnarray}
with $a = 3 + \cos (2 \alpha) $, $b = 16 \cos \alpha \cos^2 \theta$ and 
$c = \frac {\sin^4 \alpha\sin^2 (2\theta)} {8 (1 + \cos^3  \alpha \cos  \kappa \sin (2 \theta))^2} $.
%
The entanglement of formation of the same two-party state then reduces to \cite{Hill-Wootters}
\begin{equation}
E^f = H(h) \equiv -h \log_2 h - (1-h) \log_2 (1 -h),
\end{equation}
where 
\begin{equation}
 h = \frac{1 + \sqrt{1 + {\cal C}^2}}{2}.
\end{equation}
The surface of states with a vanishing discord monogamy score is then given by 
\begin{equation}
 2H(h)=H(e_1),
\end{equation}
where \(e_1\) is an eigenvalue of a single-particle density matrix of \(|\psi^s_{GHZ}\rangle\).
This surface is depicted in Fig. 1.
%
%
 \begin{figure}[h!]
\label{fig-chhobi-prothhom}
\begin{center}
\epsfig{figure=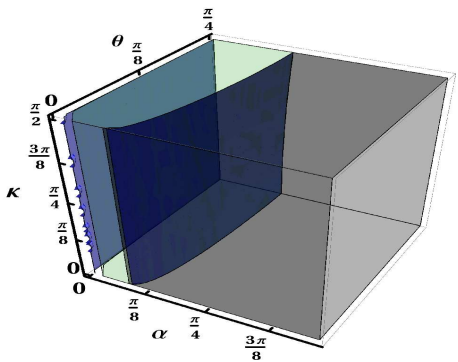, height=.2\textheight,width=0.35\textwidth}
\caption{(Color online.)  The surface in the \((\theta,\alpha,\kappa)\)-space for \(|\psi_{GHZ}^s\rangle\) with zero discord monogamy score.
Certainly, the faces \(\theta=0\) and \(\alpha=0\) of the parameter space cube have a zero discord monogamy score. However, they are not genuinely multiparty entangled. The remaining \(\delta_D=0\) state points lie on the plotted surface,
that partitions the parameter cube into two parts. 
Moreover, the state points that are within the plotted surface and the \(\alpha=0\) face have a negative discord monogamy score, while those on the other side of the plotted surface have \(\delta_D>0\). All axes are dimensionless. 
}
\end{center}
\end{figure}
To have a feel for the extent to which the states become negative and positive in the \(\delta_D \ne 0\) zones, we plot \(\delta_D\) on a path from the middle of the \(\alpha=0\) face to that of the \(\alpha=\pi/2\) face. See Fig. 2.
 \begin{figure}[h!]
\label{fig-chhobi-der}
\begin{center}
\epsfig{figure=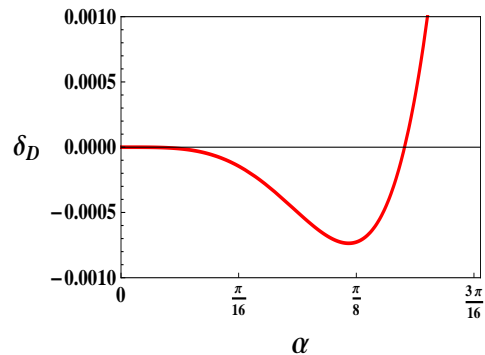, height=.2\textheight,width=0.35\textwidth}
\caption{(Color online.) Discord monogamy score against \(\alpha\) for fixed values of \(\theta\) and \(\kappa\) for the symmetric GHZ-class states. 
A straight line path beginning at \((\theta,\alpha,\kappa) = (0.4,0,1)\) and ending at \((\theta,\alpha,\kappa) = (0.4,1.57,1)\)
is considered, and the discord monogamy score is found at each point of this path. Clearly, the plotted curve must begin at the origin of the \((\alpha,\delta_D)\) plane, as the state point 
corresponds to a tri-separable state there. It then becomes negative, only to 
come back to zero, which corresponds to a point on the \(\delta_D=0\) surface plotted in Fig. 1. Subsequently, the curve enters the \(\delta_D>0\) half-plane, and remains there, to reach the ``generalized GHZ state'' \cite{GHZ} on the 
\(\alpha = \pi/2\) face.  For ease of viewing, we have cropped the \(\alpha\)-axis  at about \(\alpha = 3\pi/16\). Beyond that, the curve is monotonically increasing.  While the horizontal axis is dimensionless, the vertical 
one is measured in bits.
}
\end{center}
\end{figure}

Interestingly, unlike the W-class states, 
the GHZ-class states  with  \(\delta_D=0\) can have a nonzero genuine multipartite entanglement.  
We quantify genuine multisite entanglement by the generalized geometric measure \cite{amaderGGM}. See Fig. 3. 
\begin{figure}[h!]
\label{fig-chhobi-ditiyo}
\begin{center}
\epsfig{figure=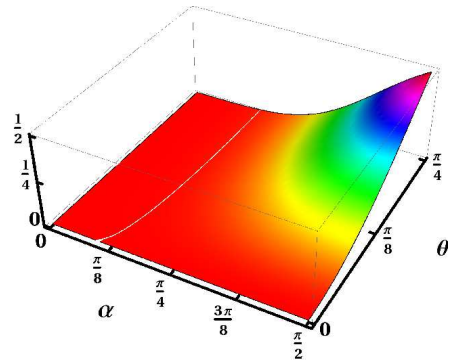, height=.2\textheight,width=0.35\textwidth}
\caption{(Color online.) GGM of symmetric GHZ-class states is plotted on the vertical axis, with respect to \(\alpha\) and \(\theta\), for \(\kappa =0.2\). The white curve lying on the GGM surface, 
depicts the state points with \(\delta_D =0\). The curve starts off with a zero GGM on the \(\alpha\) axis. However, before reaching the other end, on the \(\theta=\pi/4\) line, nonzero genuine multisite entanglement is obtained on the 
\(\delta_D=0\) curve. 
All axes are dimensionless. Although we have presented the plot for a definite value of \(\kappa\), other values of \(\kappa\) leads to qualitatively similar features, except that higher values of GGM are obtained when 
the \(\delta_D=0\) curve reaches the \(\theta=\pi/4\) line, for higher values of \(\kappa\). See Fig. 4 in this regard.}
\end{center}
\end{figure}
It is clear from Fig. 3, and by analyzing similar figures for different values of \(\kappa\), that for a given \(\kappa\), GGM is largest for \(\theta=\pi/4\) among the \(\delta_D=0\) states.  In Fig. 4, 
we plot the GGM for symmetric GHZ-class states against \(\alpha\) and \(\kappa\), for \(\theta=\pi/4\). We find that the largest value of GGM that is obtained from among the \(\delta_D=0\) states is \(\approx 0.33\). 
\begin{figure}[h!]
\label{fig-chhobi-ditiyo1}
\begin{center}
\epsfig{figure=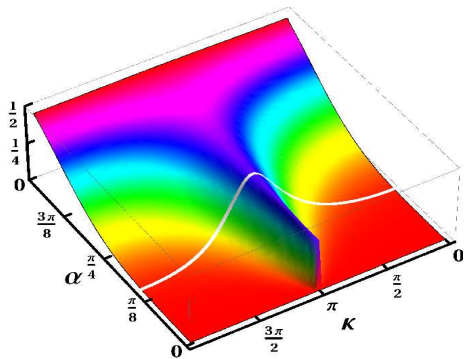, height=.2\textheight,width=0.35\textwidth}
\caption{(Color online.) GGM of symmetric GHZ-class states is plotted on the vertical axis, with respect to \(\alpha\) and \(\kappa\), for \(\theta =\pi/4\). The white curve
depicts the state points with \(\delta_D =0\). 
All axes are dimensionless. 
}
\end{center}
\end{figure}

It may also be interesting to know the status of violation of local realism of the symmetric GHZ-class states, with respect to their value of discord monogamy score. It is known that the GHZ state \cite{GHZ} maximally violates the so-called
Mermin-Klyshko Bell inequality, and therefore it is reasonable to consider this multipartite Bell inequality to look for violation of local realism of the symmetric GHZ-class states. 
%
The average of the MK operator for the symmetric GHZ-class states is given by 
\begin{eqnarray}
&& B_{MK} \equiv \mbox{tr}(B_3 |\psi_{GHZ}^{s}\rangle \langle |\psi_{GHZ}^{s}|)  \nonumber\\
&&= 4 \sin\alpha^3 \sin\theta [\cos\nu (\cos\theta \cos\kappa + \cos\alpha^3 \sin\theta)   \nonumber \\
&&+    \cos\theta \sin\nu\sin\kappa] 
\end{eqnarray}
The MK inequality will be violated if \(|B_{MK}|>1\). 
In Fig. 5, we plot the region of the parameter space in  which the Mermin-Klyshko inequality, for \(\nu=0\), is violated. We find that the region of violation is submerged in the \(\delta_D>0\) region. In particular, 
the Bell inequality employed here is not violated by the \(\delta_D=0\) states. 
However, we find that this is not a generic feature for the class of states
with  \(\delta_D=0\), as we shall see below.
\begin{figure}[h!]
\label{fig-chhobi-ditiyo2}
\begin{center}
\epsfig{figure=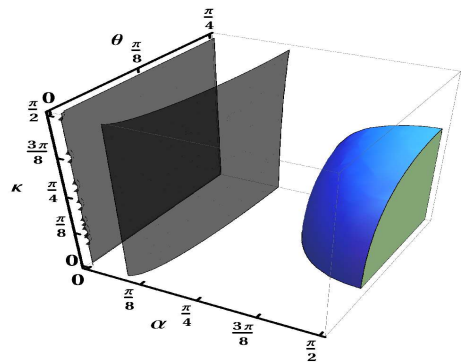, height=.2\textheight,width=0.35\textwidth}
\caption{(Color online.) Violation of Mermin-Klyshko Bell inequality for symmetric GHZ-class states. All axes are dimensionless. All depictions are as in Fig. 1, except for the octant-shaped region depicted near the \((\theta,\alpha,\kappa)=(\pi/4,\pi/2,0)\)
corner.
}
\end{center}
\end{figure}

Until now, we have been examining the set of symmetric GHZ-class states. There are however \(\delta_D=0\) states which are non-symmetric. To look into the behavior of such states, we consider a path that 
starts off from a non-symmetric GHZ-class state with a high value of GGM, but low negative value of \(\delta_D\). The path henceforth connects to the GHZ state, and is described by the following (unnormalized) state. 
%
\begin{equation}
 |\psi_{path}^{GHZ}(\mu)\rangle = \cos \mu |\phi^{GHZ}\rangle + \sin \mu |GHZ\rangle.
\end{equation}
Here, \(|\phi^{GHZ}\rangle\) is the non-symmetric GHZ-class state. It corresponds to \(\theta =0.7\), \(\kappa = 3.06\), and \((\alpha_1,\alpha_2,\alpha_3) = (0.55,0.56,0.63)\) in Eq. (\ref{radio-churi-jabar-bedanai-amar-bandhu-Jagadis}). 
And the GHZ state (unnormalized) is 
given by \(|GHZ \rangle = |000\rangle + |111\rangle \). Note that \(\mu \in [0,\pi/2]\). For \(\mu=0\), the state on the path is \(|\phi^{GHZ}\rangle\), and as mentioned before, it has a relatively high GGM, while having a low negative \(\delta_D\). 
At the other extreme, i.e. for \(\mu=\pi/2\), we have the GHZ state, which has the maximal GGM, and \(\delta_D=1\). As shown in Fig. 6, there are three values of \(\mu\) for which \(\delta_D=0\) for \(\mu \in [0,\pi/2]\). 
\begin{figure}[h!]
\label{fig-chhobi-ditiyo3}
\begin{center}
\epsfig{figure=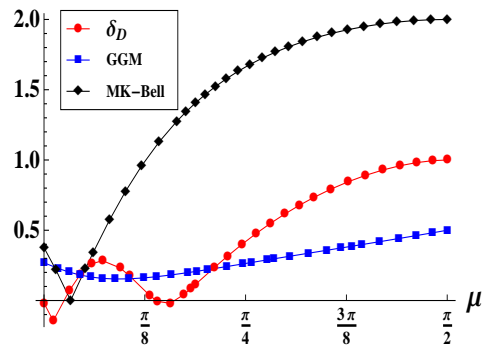, height=.2\textheight,width=0.35\textwidth}
\caption{(Color online.) 
Tracking the value of discord monogamy score as we go along the path described by  \(|\psi_{path}^{GHZ}(\mu)\rangle\). We also plot the GGM and the violation of the Mermin-Klyshko Bell inequality (for \(\nu=0\)),
 with respect to \(\mu\). The horizontal axis is dimensionless. The vertical axis is also dimensionless for the GGM and the Bell inequality violation curves, while it is in bits for \(\delta_D\). For the Bell inequality 
violation curve, the plotted quantity is \(\left|\langle \psi_{path}^{GHZ}(\mu)| B_3 |\psi_{path}^{GHZ}(\mu)\rangle\right|\). 
An interesting revelation is that the \(\delta_D=0\) states,  corresponding to the second and third crossings of the \(\delta_D\) curve with the horizontal axis, violate the MK Bell inequality.
}
\end{center}
\end{figure}

As another example of such a path, we consider the one that starts from a non-symmetric 
 W-class state that has a high value of GGM and negative \(\delta_D\). The path then again connects to the GHZ state, so that the three-party 
(unnormalized) quantum state 
describing it is given by 
\begin{equation}
 |\psi_{path}^{W \to GHZ}(\tau)\rangle = \cos \tau |\phi^W\rangle + \sin \tau |GHZ\rangle,
\end{equation}
where \(|\phi^W\rangle\) is the non-symmetric W-class state, given by 
\begin{eqnarray}
|\phi^W\rangle = 
\sin \frac{\theta_1}{2}\sin \frac{\theta_2}{2}\cos \frac{\theta_3}{2} \exp(i\phi_1)|001\rangle \nonumber \\
+\sin \frac{\theta_1}{2}\sin \frac{\theta_2}{2}\sin \frac{\theta_3}{2} \exp(i\phi_2)|010\rangle \nonumber \\
+\sin \frac{\theta_1}{2}\cos \frac{\theta_2}{2} \exp(i\phi_3)|100\rangle \nonumber \\
+\cos \frac{\theta_1}{2} |000\rangle
\end{eqnarray}
\(\theta_1 =3.25\), \(\theta_2=4.38\), \(\theta_3=11.02\), \(\phi_1=4.16\), \(\phi_2=3.98\), \(\phi_3=2.45\).
The parameter \(\tau\) lies in \([0,\pi/2]\). For \(\tau=0\), the state on the path has a relatively high GGM, and a negative \(\delta_D\), while at the other end, for \(\mu=\pi/2\), 
we have the GHZ state, which has the maximal GGM, and \(\delta_D=1\). As shown in Fig. 7, there is a unique \(\tau\) for which \(\delta_D=0\) for \(\tau \in [0,\pi/2]\). 
\begin{figure}[h!]
\label{fig-chhobi-ditiyo45}
\begin{center}
\epsfig{figure=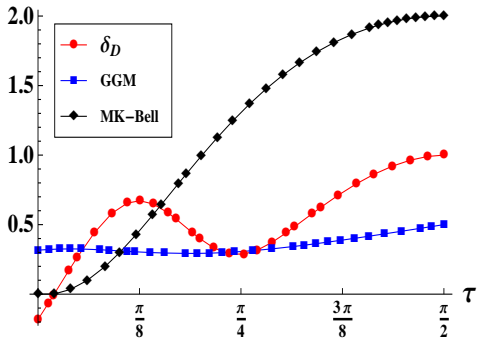, height=.2\textheight,width=0.35\textwidth}
\caption{(Color online.) Discord monogamy score and other state functions along the path described by  \(|\psi_{path}^{W \to GHZ}(\tau)\rangle\). 
The horizontal axis is dimensionless. The vertical axis is also dimensionless for the GGM and the Bell inequality violation curves, while it is in bits for \(\delta_D\). 
All other considerations, except for the chosen path, are the same as in Fig. 6.  
}
\end{center}
\end{figure}

One of the instruments that we have been using to characterize the states with a zero discord monogamy score is amount of genuine multisite entanglement that they possess, as quantified by the generalized 
geometric measure. And we have found that relatively high values of GGM can be reached by the \(\delta_D=0\) states. In Proposition III below, we show however that a maximal GGM is inaccessible to \(\delta_D=0\) states.

\noindent \textbf{Proposition III:}
For an arbitrary three-qubit pure state \(|\psi_{ABC}\rangle\),
\(\delta_D(|\psi_{ABC}\rangle) = 0 \) implies that \({\cal E}(
|\psi_{ABC}\rangle) <\frac{1}{2}\).
\\

\noindent \texttt{Proof.}
Let us suppose that there exists a pure state with
\(\delta_D(|\psi_{ABC}\rangle) = 0 \), having  \({\cal E}(
|\psi_{ABC}\rangle) =\frac{1}{2}\). That the GGM  reaches \(1/2\), implies
that the
eigenvalues of  any of the local density matrices are \(\{1/2, 1/2\}\). In particular, this implies that the state is symmetric.
Also,  the single-site as well as two-site
von Neumann entropies are of value unity. This implies that the  two-site density matrices of this three-qubit pure state are rank-2 Bell mixtures of 
two Bell states with equal probabilities.
Now an equal mixture of two Bell states is a zero-discord state. However, since the single-party von Neumann entropies are of value unity, we have \(\delta_D=1\). 
This is a contradiction. Hence, the proposition. \hfill \(\blacksquare\)

\noindent \emph{Remark.} 
We have generated $10^6 $ random three-qubit pure state points and find that the GGMs of these states with $\delta_D = 0$ are in fact bounded above by $0.35$.
Note also that the proof of Proposition III also implies that states 
with a negative discord monogamy score must have GGM \(< 1/2\). Independently, we can show that there are three-qubit pure states with a positive discord monogamy score that have maximal GGM.

\noindent \textbf{Proposition IV:}
The bipartite entanglements  of formation  of an arbitrary three-qubit pure state
\(\psi_{ABC}\rangle\) with \(\delta_D(|\psi_{ABC}\rangle) = 0 \), with \(A\) as the nodal observer, are
related to the genuine multipartite entanglement measure
\({\cal E}\) as
\begin{eqnarray}
 \label{eofvsGGM}
E^f_{AB} + E^f_{AC} \geq H({\cal E}).
\end{eqnarray}
For symmetric states, we have an equality in the above relation.
\\

\noindent \texttt{Proof.}
 \(\delta_D(|\psi_{ABC}\rangle) = 0 \implies E^f_{AB} + E^f_{AC} = S_A
= H(\lambda_A)\), where \(\lambda_A\) is the maximum eigenvalue of the 
single-site local density matrix \(\varrho_A\) of \(|\psi_{ABC}\rangle\).
Now \({\cal E} (|\psi_{ABC}\rangle) \leq 1 - \lambda_A\), and both are \(\leq 1/2\). By the well-known properties of the Shannon entropy, we have \(H(\mathcal{E}) \leq H(1-\lambda_A)= H(\lambda_A)\). Hence the relation
(\ref{eofvsGGM}). If the state is symmetric, we have \(\mathcal{E} = 1-\lambda_A\). Hence, the proposition. \hfill \(\blacksquare\)

\section{Conclusion}
\label{poune-paanch}

Summarizing, we have characterized tripartite quantum states by using monogamy properties and violation of the same of 
quantum discord. We have employed the concept of quantum monogamy score corresponding to quantum discord, which we have called discord monogamy score, for this purpose.
We have been particularly interested in the class of states having a zero quantum discord monogamy score. 
%
We found a necessary condition for vanishing
discord monogamy score for arbitrary (pure and  mixed) states in arbitrary dimensions. 
For pure states, we derived a necessary and sufficient condition. 
Specializing to the case of three-qubit pure states, multipartite
entanglement measures as well as multipartite Bell inequalities have
been used to describe different classes of states according to their
discord monogamy scores.
In particular,  we have investigated the
relation between  discord monogamy score and a genuine multipartite
entanglement measure for three-qubit pure states.

\acknowledgments
R.P. acknowledges support from the Department of Science and Technology, Government of India, in the form of an INSPIRE faculty scheme at the Harish-Chandra Research Institute, India.

\end{document}